\documentclass[a4paper]{cas-sc}

\usepackage[authoryear]{natbib}
\usepackage{graphicx} 
\usepackage{float}
\usepackage{algorithm}  
\usepackage{algpseudocode}
\usepackage{color}
\usepackage{setspace}
\usepackage[nomarkers,figuresonly]{endfloat}
\usepackage{longtable}

\def\tsc#1{\csdef{#1}{\textsc{\lowercase{#1}}\xspace}}
\tsc{WGM}
\tsc{QE}
\tsc{EP}
\tsc{PMS}
\tsc{BEC}
\tsc{DE}

\usepackage{lineno}

\begin{document}
\let\WriteBookmarks\relax
\def\floatpagepagefraction{1}
\def\textpagefraction{.001}
\shorttitle{MTH5}
\shortauthors{J. Peacock}

\title [mode = title]{MTH5: an archive and exchangeable data format for magnetotelluric time series data}

\author[1]{Jared Peacock}[type=author,auid=000,bioid=1,orcid=0000-0002-0439-0224]                  
\credit{Conceptualization, Writing--original draft, Software, Funding acquisition}

\author[2]{Karl Kappler}[type=author, auid=000,bioid=3, orcid=0000-0002-1877-1255] 
\credit{Conceptualization, Writing--original draft, Software}

\author[3]{Lindsey Heagy}[type=author, auid=000,bioid=4, orcid=0000-0002-1551-5926] 
\credit{Conceptualization, Writing--original draft, Software, Validation}

\author[4]{Timothy Ronan}[type=author, auid=000,bioid=5, orcid=0000-0001-8450-9573] 
\credit{Conceptualization, Software, Validation}

\author[5]{Anna Kelbert}[type=author, auid=000,bioid=6, orcid=0000-0003-4395-398X] 
\credit{Conceptualization, Validation, Funding acquisition}

\author[4]{Andrew Frassetto}[type=author, auid=000,bioid=2, orcid=0000-0002-8818-3731] 
\credit{Conceptualization, Writing--original draft, Funding acquisition, Supervision, Project administration}

\address[1]{U.S. Geological Survey, Geology, Minerals, Energy, and Geophysics Science Center, Moffett Field, California}
\address[2]{IMDEX Technology USA, LLC} 
\address[3]{University of British Columbia, Canada} 
\address[4]{Incorporated Research Institutions for Seismology}
\address[5]{U.S. Geological Survey, Geologic Hazards Science Center, Golden, Colorado}

\begin{abstract}
Magnetotellurics (MT) is a passive electromagnetic geophysical method that measures variations in subsurface electrical resistivity.  MT data are collected in the time domain and processed in the frequency domain to produce estimates of a transfer function representing the Earth's electrical structure.  Unfortunately, the MT community lacks metadata and data standards for time series data.  As the community grows and findability, accessibility, interoperability, and reuse of digital assets (FAIR) data principles are enforced by government and funding agencies, a standard is needed for time series data.  Presented here is a hierarchical data format (MTH5) that is logically formatted to how MT data are collected.  Open-source Python packages are also described to read, write, and manipulate MTH5 files.  These include a package to deal with metadata (\texttt{mt\symbol{95}metadata}) based on standards developed by the Working Group for Magnetotelluric Data Handling and Software assembled by the Incorporated Research Institutions for Seismology (IRIS), and \texttt{mth5}: a package to interact with MTH5 files that uses \texttt{mt\symbol{95}metadata}.  Example code and workflows are presented.            
\end{abstract}

\maketitle

\doublespacing

\section{Introduction}
\label{intro}

Magnetotellurics (MT) is an electromagnetic geophysical method that is sensitive to variations in subsurface electrical resistivity \citep{Chave2012}.  MT measurements are useful for imaging various geologic systems, including geothermal (e.g. \citet{Peacock2020}), mineral (e.g. \citet{Heinson2018}), volcanic (e.g. \citet{Bedrosian2018}), subduction (e.g. \citet{Cordell2019}), seismic (e.g. \citet{Karas2020}), and geomagnetic hazards (e.g. \citet{Kelbert2019, Murphy2021}).  Physically, MT utilizes Faraday's law of electromagnetic induction, where the Earth's magnetic field varies temporally in response to interactions of solar wind with natural magnetospheric and ionspheric current systems, inducing electric fields in the conducting Earth.  Magnetic fields are assumed to be horizontally polarized and impinge on the Earth's surface at normal incidence.  At the surface these magnetic fields diffuse into the Earth inducing electric fields. A vertical magnetic field can be induced by subsurface horizontal electrical currents.  Field measurements are collected in the time domain measuring two horizontal orthogonal electric fields, two horizontal orthogonal magnetic fields, and often a vertical magnetic field.  Subsurface electrical resistivity is estimated by transforming the time domain electromagnetic fields into the frequency domain and calculating a frequency dependent transfer function \citep{Egbert2002, Chave2004}.  This transfer function describes how the Earth's electrical structure transforms time varying magnetic fields into time varying electric fields. 

The principles of findability, accessibility, interoperability, and reuse of digital assets (FAIR) has become standard for many institutions \citep{Wilkinson2016}, as such the need for metadata and data structure standards is imperative. Unfortunately, the MT community does not have a standard for time series observations and datasets. The international MT community has traditionally been small compared to other geophysics research communities such as seismology, and MT groups are usually small clusters that have internally developed workflows to organize time series and estimate transfer functions. Sharing of the raw MT data (time series) or the processed MT data (transfer functions) has not, historically, been a priority in the MT community \citep{Kelbert2018} due to the heterogeneity of data formats and the lack of incentives for sharing.  However, the broadening of applications for MT data and increase in the adherence to FAIR principles by government and academic institutions leads us to pursue a standard format for containing MT time series data and metadata, to complement a FAIR data solution recently completed for MT transfer functions \citep{Kelbert2011, Kelbert2020}.

As part of the Seismological Facility for the Advancement of Geoscience (SAGE) facility (\url{https://www.nsf.gov/awardsearch/showAward?AWD_ID=1851048}), an National Science Foundation (NSF) sponsored effort to establish new MT science capabilities for principal investigators (PIs), the Incorporated Research Institutions for Seismology (IRIS) established a Working Group in 2019 for Magnetotelluric Data Handling and Software (\url{https://www.iris.edu/hq/about_iris/governance/mt_soft}). This group consisted of MT experts and IRIS staff, who were charged with defining a metadata standard and data format that would be used by the new MT instruments operated through IRIS, as well as support interoperability in a discipline-specific manner. The working group met remotely over two-dozen times (roughly an hour each) between spring 2019 and fall 2020 to devise, refine, and implement a framework to describe and store MT time series.  This paper presents open-source Python tools to read and write the agreed upon metadata standards \citet{Peacock2021} and data format.  Here we will describe the time series metadata, discuss the open-source Python package \verb|mt_metadata|, and introduce the data container and open-source Python package \verb|MTH5|.  All examples code in this manuscript, and more, are available as working Jupyter Notebooks in a zero-install environment using \verb|mybinder.org|.  The links are found on the GitHub landing page of each repository website by clicking the \verb|Binding| badge, or follow these links for \verb|mt_metadata|: \url{https://mybinder.org/v2/gh/kujaku11/mt_metadata/main} and \verb|mth5|:  \url{https://mybinder.org/v2/gh/kujaku11/mth5/master}.                          

\section{Time Series Metadata}
\subsection{Metadata Description}
\label{sec:attributes}

Development of time series metadata involved detailed discussion of metadata keywords, meanings, and hierarchy both within the group and with the wider MT community. To frame this approach, the group leveraged other metadata standards (e.g. Climate Forecasting (Hassell et al. 2017), International Federation of Digital Seismograph Networks (FDSN; \url{https://www.fdsn.org}), International Organization for Standardization (ISO; \url{https://www.iso.org/home.html})). Community input on the MT-specific metadata details was sought through multiple outreach efforts over the course of several years of development. Each metadata keyword is defined by 8 attributes and an example attribute (Table \ref{tab:attributes}).  The complete specification of the MT time series metadata standards can be found in \citet{Peacock2021}.  

\begin{table}
	\centering
	\caption[Metadata Attributes]{Metadata attributes}
	\begin{tabular}{>{\raggedright}p{0.9in}>{\raggedright}p{5.0in}l}
		\toprule
		\textbf{Attribute} & \textbf{Description} & \\ \midrule
		\textbf{Name} & A full descriptive name that is logical for the keyword.  Forced to be all lower case and full words separated by \verb|_| & \\
		\textbf{Type} & Base data type  [ string | float | int | Boolean ] & \\
		\textbf{Style} & Describes how the string should be formatted.  Options are included in Table \ref{tab:strings} & \\ 
		\textbf{Required} & \verb|True| if required or \verb|False| if optional &  \\
		\textbf{Units} & Physical units of the keyword. Given as full name all lowercase separated by \verb|_| and a \verb|-| for multiplicative units (e.g. ohm-meters) and \verb|per| for a ratio of units (e.g. meters per second) & \\
		\textbf{Description} & Detailed description of what the keyword represents & \\
		\textbf{Options} & If "style" is "controlled vocabulary" this is a list of accepted options & \\
		\textbf{Example} & An example use of the keyword &\\ 
		\textbf{Default} & A default value, only set if Required = True &\\ \bottomrule

	\end{tabular}
	\label{tab:attributes}
\end{table}

\begin{table}
	\centering
	\caption[Acceptable String Formats]{Acceptable String Formats}
	\begin{tabular}{>{\raggedright}p{.9in}>{\raggedright}p{3.2in}c}
		\toprule
		\textbf{Style} & \textbf{Description}  & \textbf{Example} \\ \midrule
		Free Form & An unregulated string that can contain \{a-z, A-Z, 0-9\} and special characters & This is Free Form! \\ 
		Alpha Numeric & A string that contains no spaces and only characters \{a-z, A-Z, 0-9, -, \_\} & WGS84 \\
		Controlled Vocabulary & Only certain names or words are allowed. In this case, examples of acceptable values are provided in the documentation as [ option01 $|$ option02 $|$ ... ]. The ... indicates that other options are possible but have not been defined in the standards yet &  reference\_frame = geographic \\ 
		List & List of entries using a comma separator & Ex, Ey, Hx, Hy, Hz, T \\
		Number & A number according to the data type; number of decimal places has not been implemented yet & 10.0 (float) or 10 (integer) \\ 
		Date & ISO formatted date YYYY-MM-DD in UTC & 2020-02-02 \\
		Date Time & ISO formatted date time YYYY-MM-DDThh:mm:ss.ms+00:00 in UTC & 2020-02-02T12:20:45.123456+00:00 \\ 
		Email & A valid email address & \url{person@mt.org} \\
		URL & A full URL that a user can view in a web browser  &  \url{https://www.passcal.nmt.edu/} \\ \bottomrule
	\end{tabular}
	\label{tab:strings}
\end{table}

\subsection{Metadata Hierarchy}
\label{sec:structure}

Metadata keywords are structured to be hierarchical such that more complicated metadata can be represented as a combination of simpler keywords.  This is done by combining keywords with a period to create nested keywords, for example to describe the latitude of a location one could use \verb|location.latitude|.  

\begin{figure}
	\includegraphics[width=\textwidth]{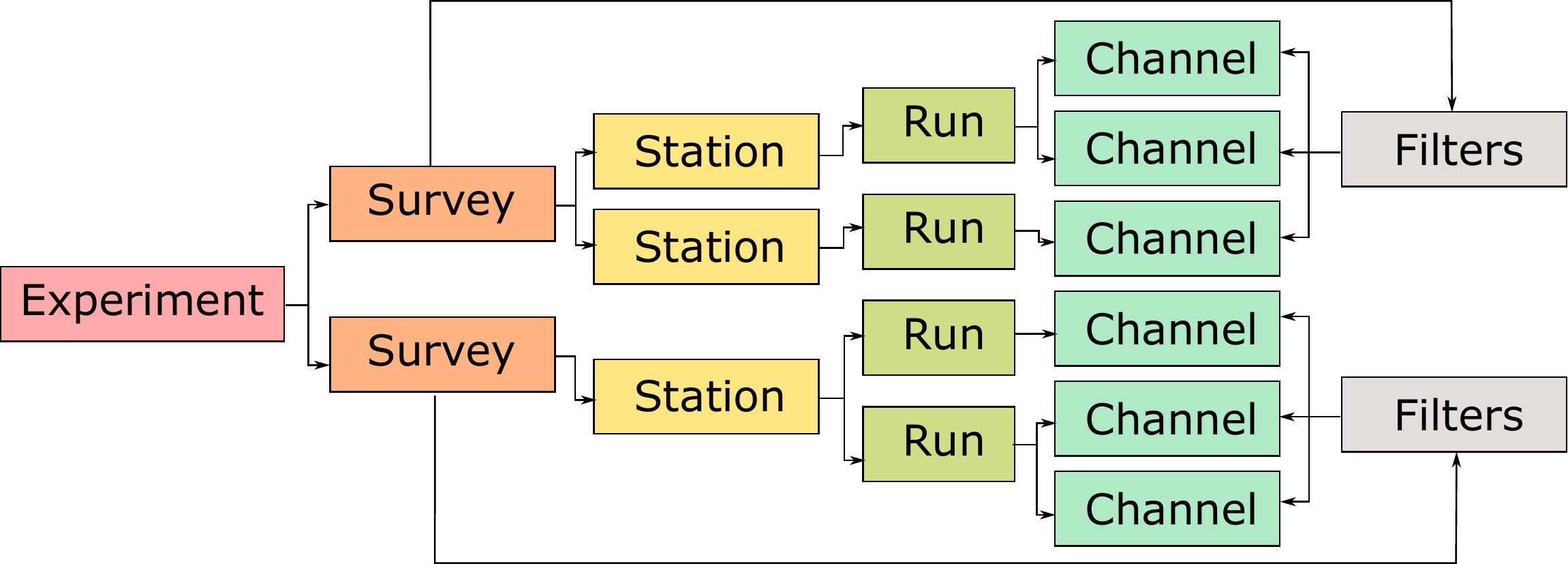}
	\caption{Example of how MT time series metadata are structured.  Top level is an \texttt{Experiment}, then \texttt{Survey} which contains all the filter metadata for the survey as a dictionary keyed by filter name.  The next level down is \texttt{Station}, then \texttt{Run}, and finally \texttt{Channel}.  The \texttt{Channel} metadata contains the names of filters used and whether they have been applied.  To get filter metadata, retrieve the metadata from the filter dictionary stored at the \texttt{Survey} level.} 
	\label{fig:metadata}
\end{figure}

The hierarchy and structure of the MT metadata logically follows how MT time series data are collected (Figure \ref{fig:metadata}). The highest level is an \verb|Experiment| which contains all metadata for an MT experiment collected in a geographical region. The next level down is \verb|Survey| which contains metadata for data collected over a certain time interval in a given geographic region. This may include multiple PIs or data collection episodes but should be confined to a specific project. Next level down is a \verb|Station| which contains metadata for a single location over a certain time interval. Beneath station is the \verb|Run| level.  A \verb|Run| contains metadata for continuous data collected at a single sample rate.  If the station location changes during a run, then a new \verb|Station| should be created and subsequently a new \verb|Run| under the new station. If the sensors, cables, data logger, battery, etc. are replaced during a run but the station remains in the same location, then this can be recorded in the \verb|Run| metadata but does not require a new station. Finally, a \verb|Channel| contains metadata for a single channel during a single run, where electric, magnetic, and auxiliary channels have specific metadata to uniquely describe the physical measurement. Metadata standards for each level are defined in \citet{Peacock2021}. If channel parameters are changed between runs, this would require creating a new run.  If a run has channels that drop out, the start and end period will be the minimum time and maximum time for all channels recorded and those channels that dropped will be filled with null values.  

A \verb|Filters| level exists at the \verb|Survey| level and describes all filters applied to the time series data to get calibrated data in useful physical units.  The \verb|Filters| level is placed here to remove redundancy at the channel level of repeating filter metadata.  Currently five different types of filters are supported: frequency-amplitude-phase tables, pole-zero filters, finite impulse response filters, coefficient filters, and time-delay filters.  Each channel has two keywords that define the filters and both are lists of the same length: \verb|channel.filter.name| is a list of filter names to calibrate the data and \verb|channel.filter.applied| is a list of booleans indicating whether the filter has been applied (true) or not (false).  Both of these lists must be ordered the same as how the filters are applied.  An applied filter means the data have been transformed in some way, for example calibrating the data from physical units to digital counts.    

\subsection{Formatting Standards}

Specific and required formatting standards for location, time and date, and angles are defined to follow international standards and MT conventions.

\subsubsection{Time and Date Format}

All time and dates are given as an ISO formatted date-time string in the coordinated universal time (UTC) time zone.  The ISO date-time format is \verb|YYYY-MM-DDThh:mm:ss.ms+00:00|, where the UTC time zone is represented by \verb|+00:00|. UTC can also be denoted by \verb|Z| at the end of the date-time string \verb|YYYY-MM-DDThh:mm:ss.msZ|.  Note that \verb|Z| can also represent Greenwich Mean Time (GMT) but is an acceptable representation of UTC time.  If the data requires a different time zone, this can be accommodated, however UTC is preferred whenever possible to avoid confusion of local time and local daylight savings. Milliseconds can be accurate to nine decimal places.  ISO dates are formatted \verb|YYYY-MM-DD|, and ISO hours are given as a 24 hour number or military time, e.g. 4:00 AM is 04:00 and 4:00 PM is 16:00.

\subsubsection{Location}

All latitude and longitude locations are given in decimal degrees in the datum specified at the \verb|Survey| level. The datum is static, and should follow the well-known text format described by the Open Geospatial Consortium \citep{OGC2019}, for example WGS84.  The entire survey should use only one datum that is specified at the Survey level, therefore the user must project all coordinates to the specified datum. All locations are relative to the prime meridian (0, 0).  Latitude values must be within $[-90, 90]$, where negative values represent locations south of the prime meridian.  Longitude values must be within $[-180, 180]$, where negative values are west of the prime meridian.  Elevation and other distance values are given in meters.  The preferred datum is WGS84 but others are acceptable.

\subsubsection{Angles}

\begin{figure}
	\centering
	\includegraphics[width=.85\textwidth]{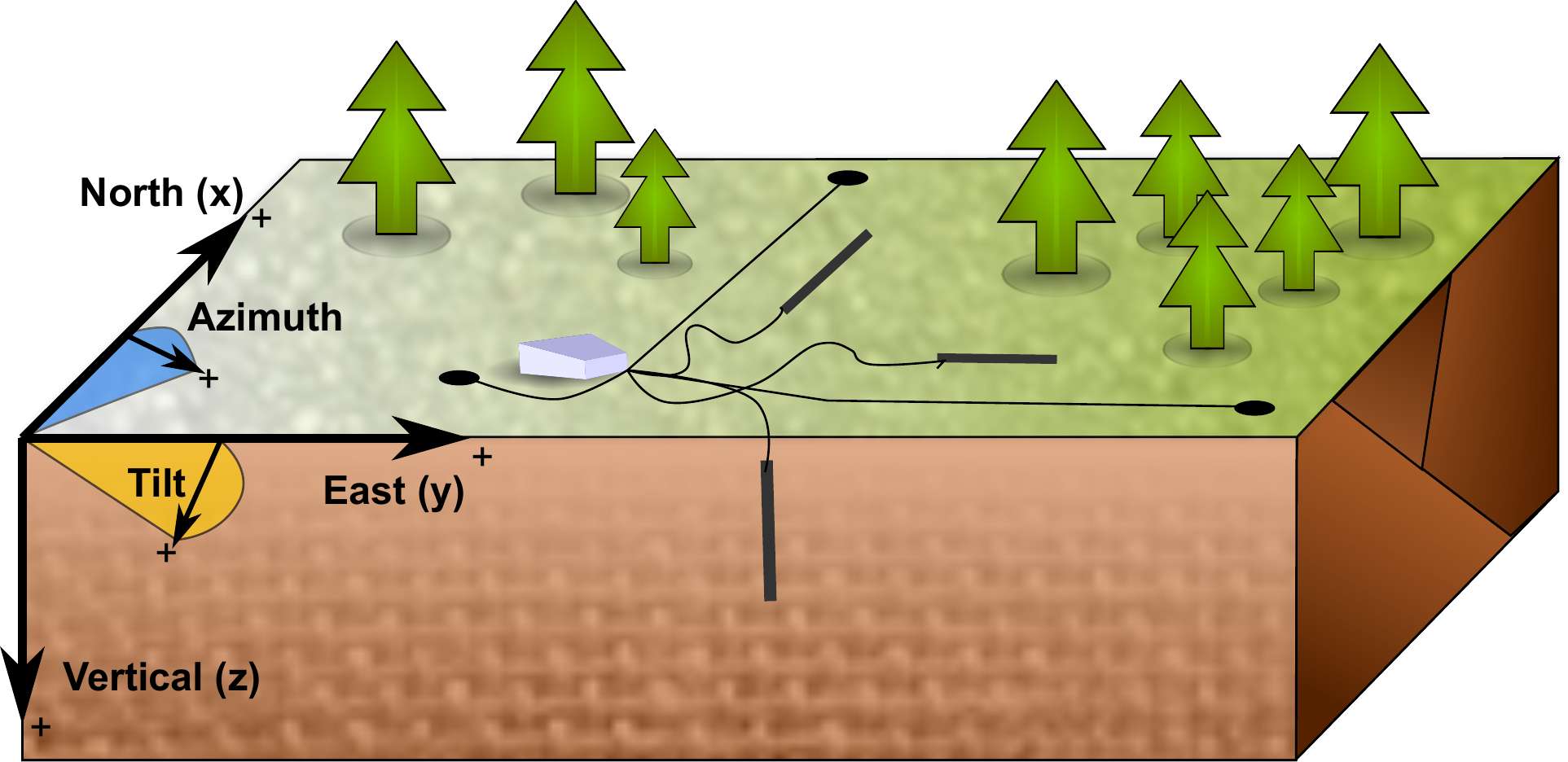}
	\caption{Diagram showing a right-handed geographic coordinate system.  The azimuth is measured positive clockwise along the horizontal axis and tilt is measured from the horizontal axis with positive down = 90, positive up = 180, and horizontal = 0.}
	\label{fig:reference}
\end{figure}

All angles of orientation are given in decimal degrees.  Orientation of channels should be given in a geographic or a geomagnetic reference frame where the right-hand coordinates are assumed to be north = 0, east = 90, and vertical is positive downward (Figure \ref{fig:reference}).  The coordinate reference frame is given at the station level \\ \verb|station.orientation.reference_frame|.  Two angles to describe orientation of a sensor are given by \\ \verb|channel.measurement_azimuth| and \verb|channel.measurement_tilt|.  In a geographic or geomagnetic reference frame, the azimuth refers to the horizontal angle relative to north positive clockwise, and the tilt refers to the vertical angle with respect to the horizontal plane. In this reference frame, a tilt angle of 90 points downward, 0 is parallel with the surface, and -90 points upwards.  

Archived data should remain in measurement coordinates. Any transformation of coordinates for derived products can be stored in the transformation angles at the channel level in \verb|channel.transformed_azimuth| and \\ \verb|channel.transformed_tilt|, and the transformed reference frame can then be recorded in \\ \verb|station.orientation.transformed_reference_frame|.      

\subsection{Units}

Acceptable units are only those from the International System of Units (SI).  Only long names in all lower case are acceptable.  Units with multiple dimensions should be separated by a \verb|-| if multiplicative, or \verb|per| if divided.  For example velocity would be \verb|meters per second| and resistivity would be \verb|ohm-meter|.     

\subsection{Open-source Python package: mt\_metadata}

To help users handle, validate, and manipulate MT time series metadata, an open-source Python package has been developed: \verb|mt_metadata|.   The base container provides functions to read/write XML, JSON, and Python dictionaries, and validates each metadata keyword and value against attributes described in Section \ref{sec:attributes}.  This base container is inherited by containers representing each level of MT metadata.  All time series metadata objects are included in the \verb|mt_metadata.timeseries| module.  

The internal structure of the base class begins with reading in the metadata standards defined by \citet{Peacock2021}, which are stored as JSON files included in the distribution.  The standards files are read in as a Python dictionary and stored as a private variable.  The keys of this private dictionary are attribute names and values are dictionaries describing the metadata attribute (Table \ref{tab:attributes}). The base class is also assigned attributes that are the key names for the user to access.   When the value of a base class attribute is modified, the value is validated against the defined standards in the following order: "name", "data type", "style", and "options" using validation functions for each standard type.  If the input data is not the required data type, the data type validator will attempt to convert the value to the required data type.  If validation fails at any point, an error is raised.  Note that the base class is not limited to MT data, but any standardized metadata following Section \ref{sec:attributes}.

\begin{figure}
	\includegraphics[width=.75\textwidth]{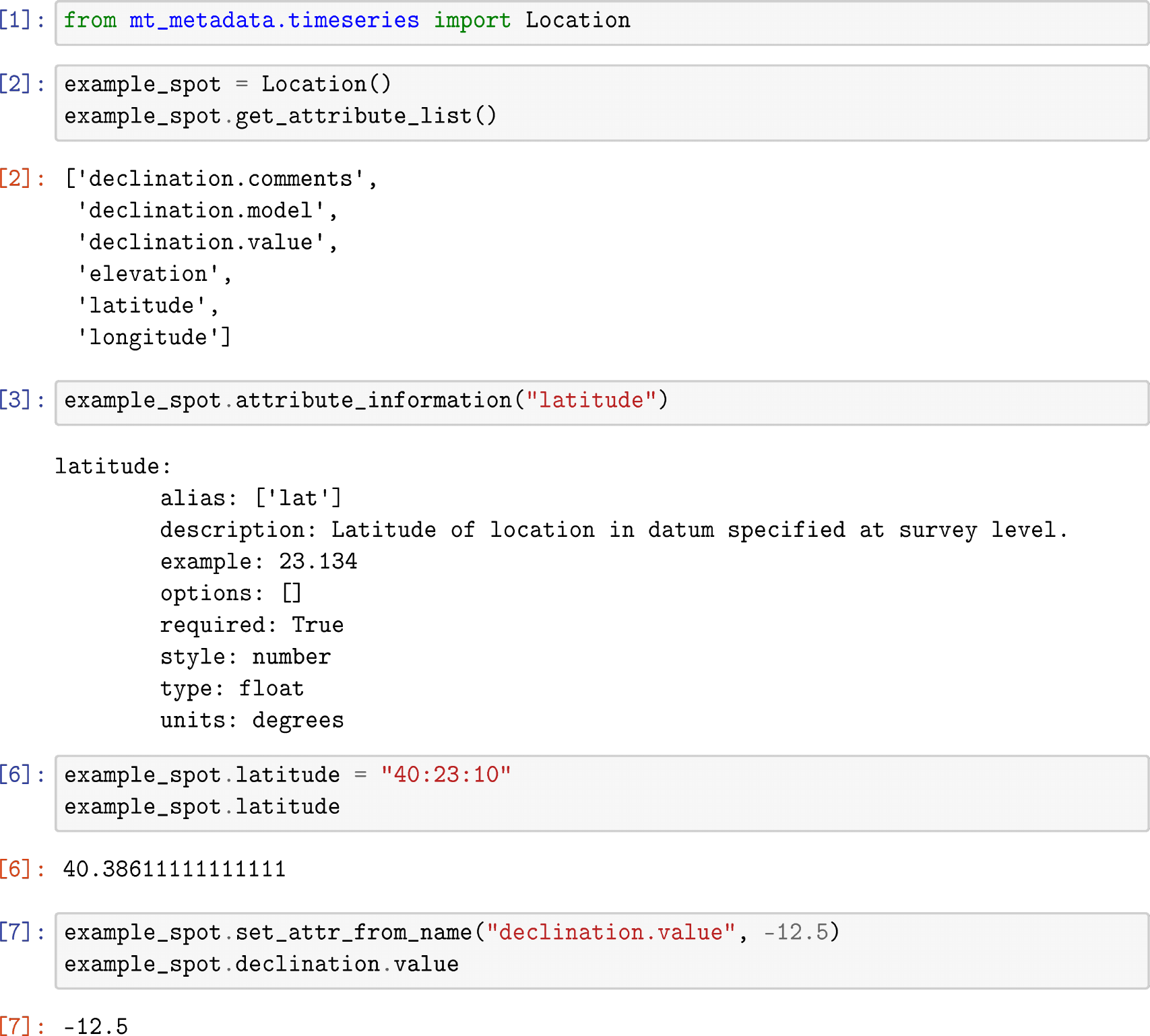}
	\caption{Example code for how to access the attribute list of a metadata class, get a description of an attribute, and how the validators convert input to the appropriate data type. Source code can be found at \url{https://github.com/kujaku11/mt_metadata/tree/main/examples/notebooks/example_01.ipynb}}
	\label{fig:code01}
\end{figure}

The base class provides methods to get information about attributes, add new attributes, get and set attributes, print string representation, read/write XML, JSON, and Python dictionaries, update from a dictionary, and compare similar objects.  Example Jupyter Notebook code is provided in Figure \ref{fig:code01}.

\begin{figure}
	\includegraphics[width=.5\textwidth]{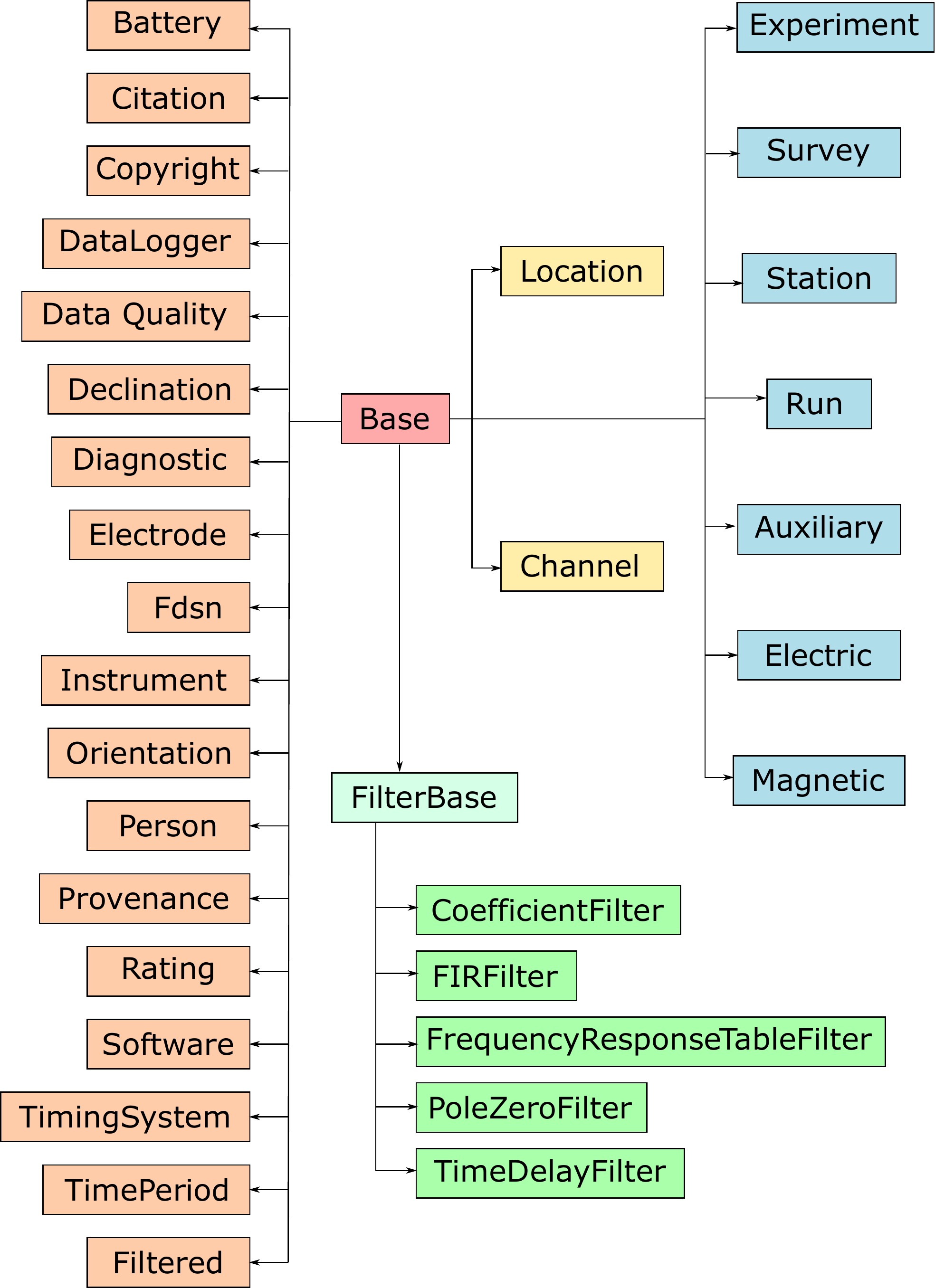}
	\caption{Inheritance plot of the Base class for time series metadata objects.  Objects on the left are used as attributes from more complex objects on the right.  For example "survey.provenance.author".  The filters have a base class that inherits the main base class.  FIRFilter: finite impulse response filter; FDSN: International Federation of Digital Seismograph Networks.}
	\label{fig:inheritance}
\end{figure}

Time series metadata have been broken down into the primitive representation and made into class objects that inherit the base class (Figure \ref{fig:inheritance}).  These are then used as attributes for more complex objects.  For example \verb|declination| describes how declination was estimated, which can be an attribute of \verb|location|. The declination value for the given location is then \verb|location.declination.value| (Figure \ref{fig:code01}).    

\begin{figure}
	\includegraphics[width=.75\textwidth]{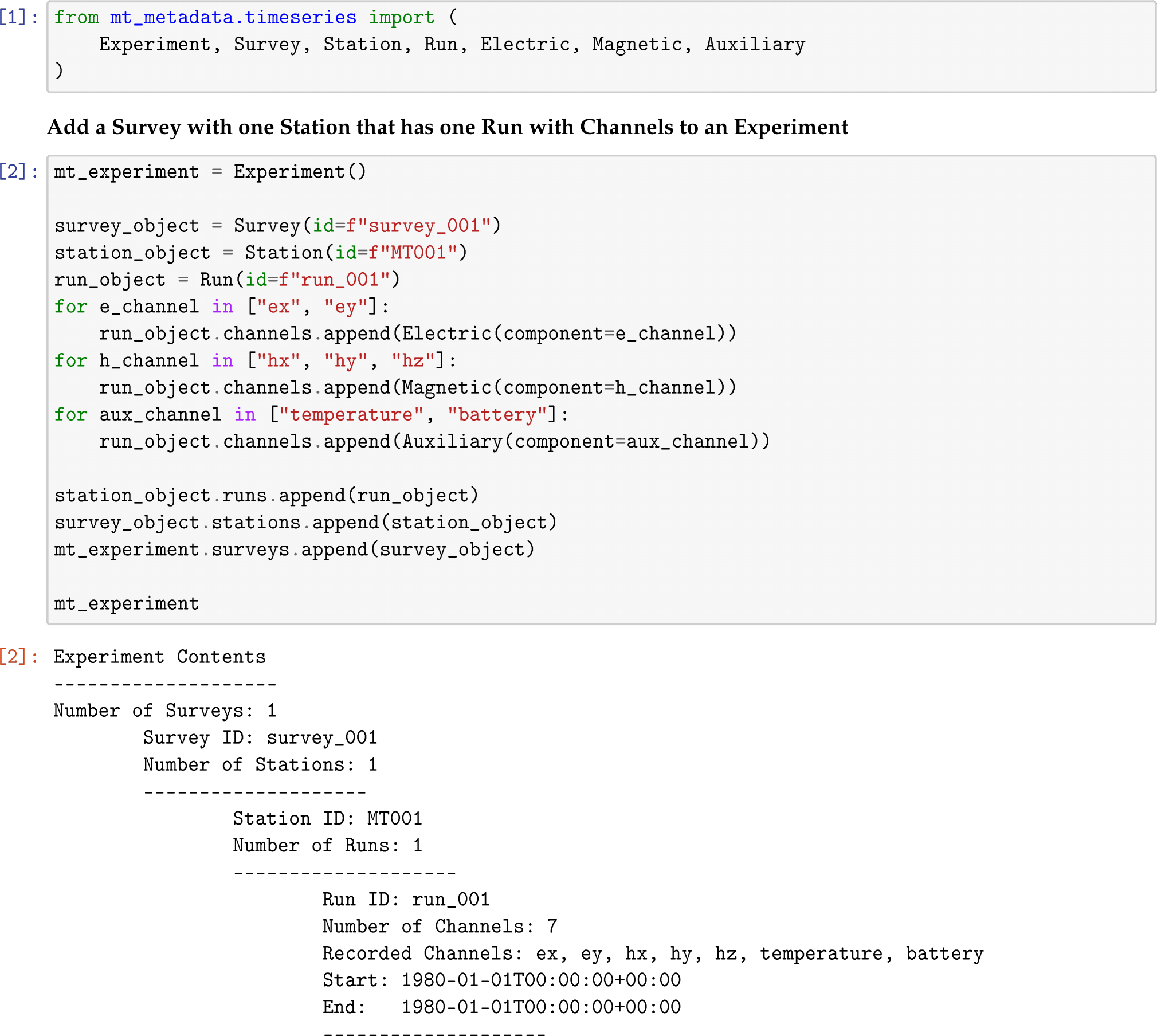}
	\caption{Example code of how to make an \texttt{Experiment} metadata object from scratch and examine its contents. Source code can be found at \url{https://github.com/kujaku11/mt_metadata/tree/main/examples/notebooks/example_02.ipynb}.}
	\label{fig:code04}
\end{figure}

Following the structure displayed in Figure \ref{fig:metadata}, each level container includes an attribute that is a list of the containers for the next level down (Figure \ref{fig:code04}).  For example, \verb|Experiment| has an attribute \verb|Experiment.surveys| which is a list of \verb|Survey| objects.  Moreover, \verb|Survey| has an attribute \verb|Survey.stations|, which is a list of \verb|Station| objects, and so on down the chain to \verb|Channel|.  This provides a logical structure of metadata objects and allows for writing and reading comprehensive metadata. More information can be found at \url{https://mt-metadata.readthedocs.io/en/latest/}.   

\section{MTH5}
\label{sec:mth5}

Several geophysical communities format data in hierarchical data formats like HDF5 and NetCDF-4 because these formats contain data and metadata in the same place for a self describing logically structured data set.  HDF and HDF5 files have mutliple benefits \citep{HDFGroup2019}:  1) the base code is open-source and community driven;  2) files are flexible;  3) file size is only limited by available resources; 4) files are portable across nearly all operating systems and platforms from laptops to cloud based parallel systems, and have no limitations on the number of data objects contained within the file ;  5) RAM requirements are optimized with a high-performance input/output system to only load requested data;  6) chunking and compression are inherent to optimize efficient storage and retrieval.  A single writer with multiple readers (SWMR) is supported, but parallel reading/writing needs parallel HDF5 (PHDF5).  For cloud environments the HDF Group provides highly scalable data services (HSDS).     

Most data collected by satellites are stored as platform specific HDF5 files that provide users with multi-channel map-oriented products (e.g. \cite{Lee2012, Klein2016, Loomis2019}).  For ground based measurements, an adaptable seismic data format (ASDF) has been developed to store various seismic data with provenance \citep{Krischer2016}.  The authors suggest ASDF could be extended to other geophysical data types in the future.  IRIS Portable Array Seismic Studies of the Continental Lithosphere (PASSCAL) also developed an HDF5 container the PASSCAL hierarchical format (PH5), which is their contemporary format to archive seismic data for large assembled datasets.  We pursued extending ASDF and PH5 but decided to build a specific container for MT data (MTH5) that could potentially be an extension in the future to ASDF, PH5, and the planned geophysical format being developed by IRIS and UNAVCO, which will be a flexible format to store various types of geophysical data \citep{Habermann2021}. MTH5 is an HDF5 file and the open-source Python module \verb|mth5| provides tools to interface with the file.

\begin{figure}
	\includegraphics[width=.9\textwidth]{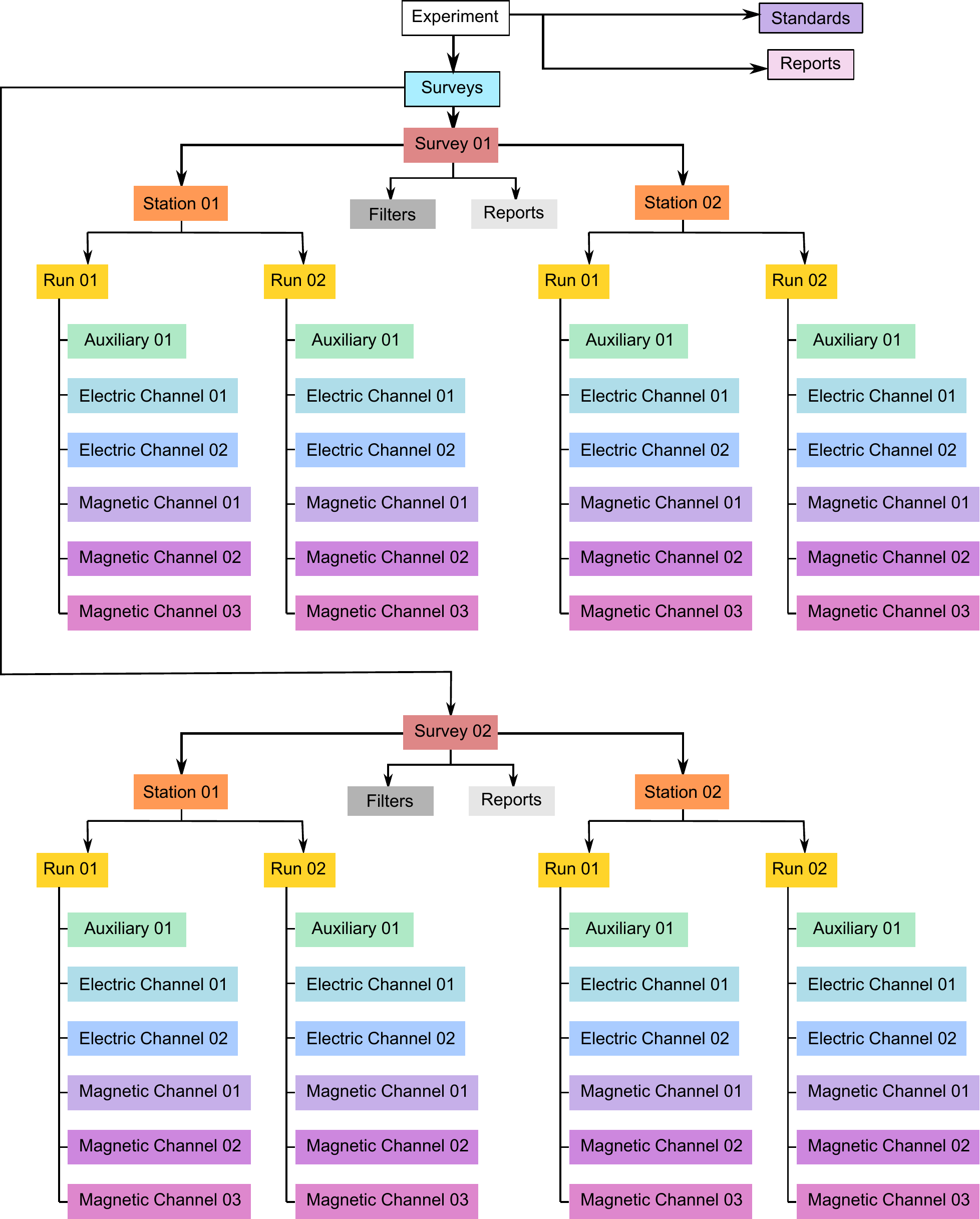}
	\caption{Schematic of the MTH5 file structure.  The top group is \texttt{Experiment} which has three subgroups: \texttt{Surveys}, \texttt{Reports} and \texttt{Standards}.  The \texttt{Standards} group contains the metadata standards used to make the file and is stored as a table for easy reference.  The \texttt{Reports} group can contain any auxiliary reports that compliment or help describe the data.  The \texttt{Surveys} group can contain multiple \texttt{Survey} groups.  A \texttt{Survey} group contains a \texttt{Filters} group, which has a subgroup for each supported filter (see Figure \ref{fig:inheritance} for supported filters), and contains all filters for a given survey.  A \texttt{Survey} group can have multiple \texttt{Station} groups.  A \texttt{Station} group can have multiple \texttt{Run} groups.  A \texttt{Run} can have multiple \texttt{Channel} data sets which is ultimately where data are stored.  Each group and data set includes metadata attributes of the appropriate level.  Each \texttt{Survey}, \texttt{Station}, \texttt{Run} group is named by its \texttt{id}, and \texttt{Channel} data sets are named by their \texttt{component}.  For example the path to \texttt{MagneticChannel02} of \texttt{Survey02} would be \texttt{Experiment/Survey\_02/Station\_01/Run\_02/Magnetic\_Channel\_02}.}
	\label{fig:mth5}
\end{figure}       

Two main types of structures exist in HDF5 files: groups and data sets.   Groups are similar to folders in a filing system but with the ability to store metadata as attributes.  Data sets store data and can also have metadata attached.  The structure of an MTH5 file follows how MT time series data are collected (Figure \ref{fig:mth5}).   Metadata follows the standards described in \citet{Peacock2021} and are parallel to the MTH5 structure.  The root group is \verb|Experiment|, with metadata \verb|experiment|, which has three subgroups: \verb|Surveys|, \verb|Reports|, and \verb|Standards|.  The \verb|Surveys| group holds multiple \verb|Survey| groups that are named by their \verb|survey.id|. The \verb|Reports| group can contain auxiliary files that provide additional context for the data, like a field report. Currently, it is up to the user to get the auxiliary file into a serializable format that can be stored in an HDF5, like an image or a portable document format (PDF) report.  Finally, the \verb|Standards| group contains an array of metadata standards used for the file. The \verb|Standards| group contains a data set with a column for each row in Table \ref{tab:attributes} and each row describes a metadata keyword.  This is done to allow users to know the standards in-place without having to look them up.   

A \verb|Survey| group contains three subgroups: \verb|Stations|, \verb|Filters|, and \verb|Reports|.  The \verb|Stations| group contains the \verb|Station| groups named by their \verb|station.id|.  Each \verb|Station| group has associated \verb|station| metadata and contains \verb|run| groups named by their \verb|run.id|.  A run group has \verb|run| metadata and contains \verb|Channel| data sets named by their \verb|channel.component|.  Each channel has its own group and appropriate metadata.  Currently the accepted channels are electric, magnetic, and auxiliary.  Auxiliary can be anything that is not an electric or magnetic channel, the only requirement is that it has the same sample rate as the other channels.  

The \verb|Filters| group in the \verb|Survey| group contains groups for nearly all filters typically encountered in an MT survey.  The \verb|zpk| group contains a pole-zero type filter where the poles and zeros are stored as data sets with a complex number data type.  The \verb|fap| group contains frequency-amplitude-phase lookup tables, which are stored as an array of N~x~3 with a column for each of N frequency, amplitude, and phase entries.  The \verb|coefficient| group contains coefficient type filters where the data are multiplied by a single number to convert units or scale the data properly.  The \verb|time_delay| group contains time-delay filters where the data is time shifted by a single number, common in digitizers with multiple channels.  The \verb|fir| group contains finite impulse response filters where the coefficients are stored as a N~x~1 array.

Each MTH5 file has attributes that describe the file.  These are defined in Table \ref{tab:mth5_attributes}.  These help readers understand how to read the file properly, for example making sure the structure matches the given file version.

\begin{table}
	\centering
	\caption[MTH5 File Attributes]{MTH5 File Attributes}
	\begin{tabular}{>{\raggedright}p{1.5in}>{\raggedright}p{4.4in}l}
		\toprule
		\textbf{Attribute} & \textbf{Description} & \\ \midrule
		\texttt{file.type} & Type of file will always be \texttt{MTH5} for an MTH5 file, but provided for future flexibility  & \\
		\texttt{file.version} & MTH5 file version, currently 0.2.0 & \\
		\texttt{file.access.platform} & The operating system used to create the file and read the file. & \\ 
		\texttt{file.access.time} & The time the file was last accessed YYYY-MM-DDThh:mm:ss+00:00 &  \\
		\texttt{mth5.software.version} & Version of the software used to make the file & \\
		\texttt{mth5.software.name} & Name of the software used to make the file & \\
		\texttt{data\_level} & Data level contained in the file.  0 = raw data + metadata derived from the data logger; 1=raw data + full metadata; 2=derived product (e.g. conversion to physical units) &   \\ \bottomrule
		
	\end{tabular}
	\label{tab:mth5_attributes}
\end{table}

\subsection{Open-source Python package: mth5}

To read/write, manipulate, validate, and interact with an MTH5 file, the open-source Python package \verb|mth5| has been developed, which leverages \verb|h5py| to interact with HDF5 and \verb|xarray| \citep{Hoyer2017, Hoyer2021} to store time series data in a usable way. \verb|h5py| was chosen because data stored are mainly one-dimensional arrays as opposed to \verb|pytables| which is better for storing multi-dimensional tabular data.  Benefits of \verb|xarray| include native objects that contain data and metadata, lazy access, multi-dimensional indexing of data sets, efficient data frame tools, and native conversion to Dask and Zarr arrays for parallel computing. 

Each group described in Section \ref{sec:mth5} has a corresponding object in \verb|mth5| which inherits a \verb|BaseGroup| object that is a convenience class to access an HDF5 group.  The main role of \verb|BaseGroup| is to validate metadata against the standards by using the appropriate \verb|mt_metadata.timeseries| metadata object.  \verb|BaseGroup| also provides convenience methods to list all contained subgroups and initiate a group with the appropriate metadata, or if a group already exists read the metadata into the appropriate \verb|mt_metadata.timeseries| metadata object.  All references in \verb|BaseGroup| to an HDF5 group are weak, meaning the objects created from the HDF5 file are not protected from Python's garbage collector and will be removed when the file is closed. 

Each group object contains methods to add, get, and remove a group at the next level down.  For example the object containing a station group has methods to add, get, and remove a run.  Note that removing a group in HDF5 is not like deleting the group, it is just removing the reference to that group.  Data sets also have a base object in \verb|mth5|.  Similar to \verb|BaseGroup|, \verb|ChannelDataset| validates metadata, and contains methods to output data into common Python objects like a \verb|numpy.ndarray| \citep{Harris2020}, \verb|pandas.DataFrame| \citep{McKinney2010, Reback2021}, \verb|xarray.DataArray|, and a \verb|mth5.timeseries.ChannelTS| object.  The benefit of the last three is that the time series is indexed by time for easy indexing and slicing.  The \verb|ChannelTS| object is a wrapper to interact with an \verb|xarray.DataArray| object that contains the data and validates metadata through the appropriate \verb|mt_metadata.timeseries| metadata object.      

The \verb|Run| group is a bit more complicated because it contains the channels as individual data sets. A run is defined here as a collection of synchronous channels recorded at the same sample rate.  \verb|RunTS| is an object that provides the user with a run container where all channels in a run are contained in one object. \verb|RunTS| is a wrapper for an \verb|xarray.DataSet| which is a collection of \verb|xarray.DataArray| objects.  A \verb|RunTS| object is indexed by time where all channels are aligned by the earliest start time and the latest end time, and misalignment is filled with null values (NaNs).  Metadata for each channel is maintained and accessible.  A \verb|RunTS| is meant to be the main object used for time series processing, where a processing run would be a collection of data collection runs.  

The main class for users is the \verb|MTH5| module which provides methods to open and close MTH5 files, add, get, and remove surveys, stations, runs, and channels, and includes groups directly under \verb|Experiment|.  In the \verb|Surveys| group, a property that summarizes all channels in a given MTH5 file is provided.  This summary table is a \verb|pandas.DataFrame| which provides the user with a searchable object to locate data, and is directly derived from the data and metadata.  A column in the summary table represents an HDF5 reference to the channel which can be used to directly get the data for that channel.  From the summary table, a user can search for all channels recorded during a given time segment to do multi-station processing.   The data can be accessed from all the HDF5 references for channels returned in the query.  More information can be found at \url{https://mth5.readthedocs.io/en/latest/}.

\subsubsection{Building an MTH5 File}
\label{sec:build}

There are many ways a MTH5 file can be built with \verb|mth5|.  The most basic is to manually input metadata and data from a Python shell, but this should only be used for small changes.  Most users will have data from a data logger, however each data logger may produce files in different formats with varying levels of metadata.  For flexibility, a plug-in architecture is developed in \verb|mth5|, where a plug-in reader can be built for specific data files, similar to ObsPy \citep{Krischer2015}.  The only requirement is that the \verb|read_{data_logger}| function in the plug-in must output a \verb|RunTS| object with the appropriate metadata.  For example, if the file type is made by an "example" data logger and the output files are \verb|.dat| files, then a reader module would be developed with the name \verb|example| that has a function \verb|read_example|.  This module would be added to the \verb|mth5.io| module. The function and file types would then be added to the dictionary of available functions and associated file types that the main \verb|read| function references.  From the \verb|RunTS| object, a station, run, and channel data sets can be added to an MTH5 file.  Currently, there are plug-ins for Zonge International Z3D files, U.S. Geological Survey ASCII files, NIMS BIN files, LEMI424 long period ASCII files, and miniSEED files via ObsPy.

\begin{figure}
	\includegraphics[width=.75\textwidth]{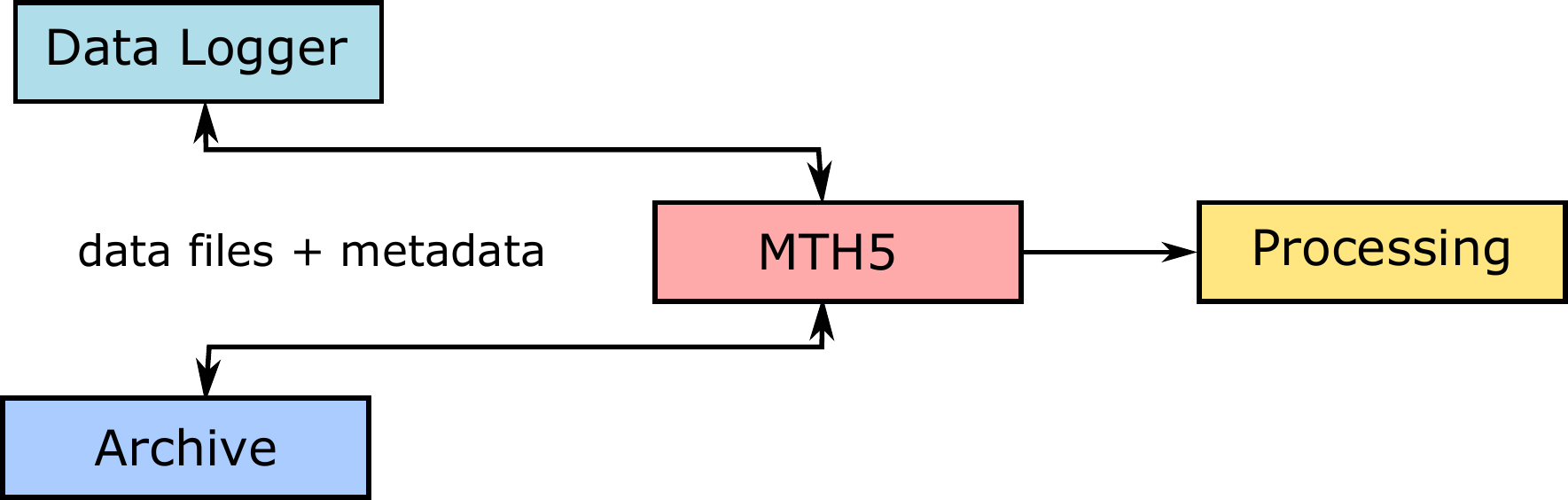}
	\caption{Example workflow for using a MTH5 file.  A MTH5 file can be an archive format and a working format for processing.  An MTH5 can be built from an archive or by reading from a data logger and adding metadata not included in the raw files.  How data are transferred is archive dependent.  For example, the IRIS DMC will send and receive miniSEED data files and StationXML metadata files. Note: tools exist in \texttt{mth5} and \texttt{mt\symbol{95}metadata} to convert an MTH5 to and from miniSEED and StationXML.}
	\label{fig:workflow}
\end{figure}           
                           
\begin{figure}
	\includegraphics[width=.75\textwidth]{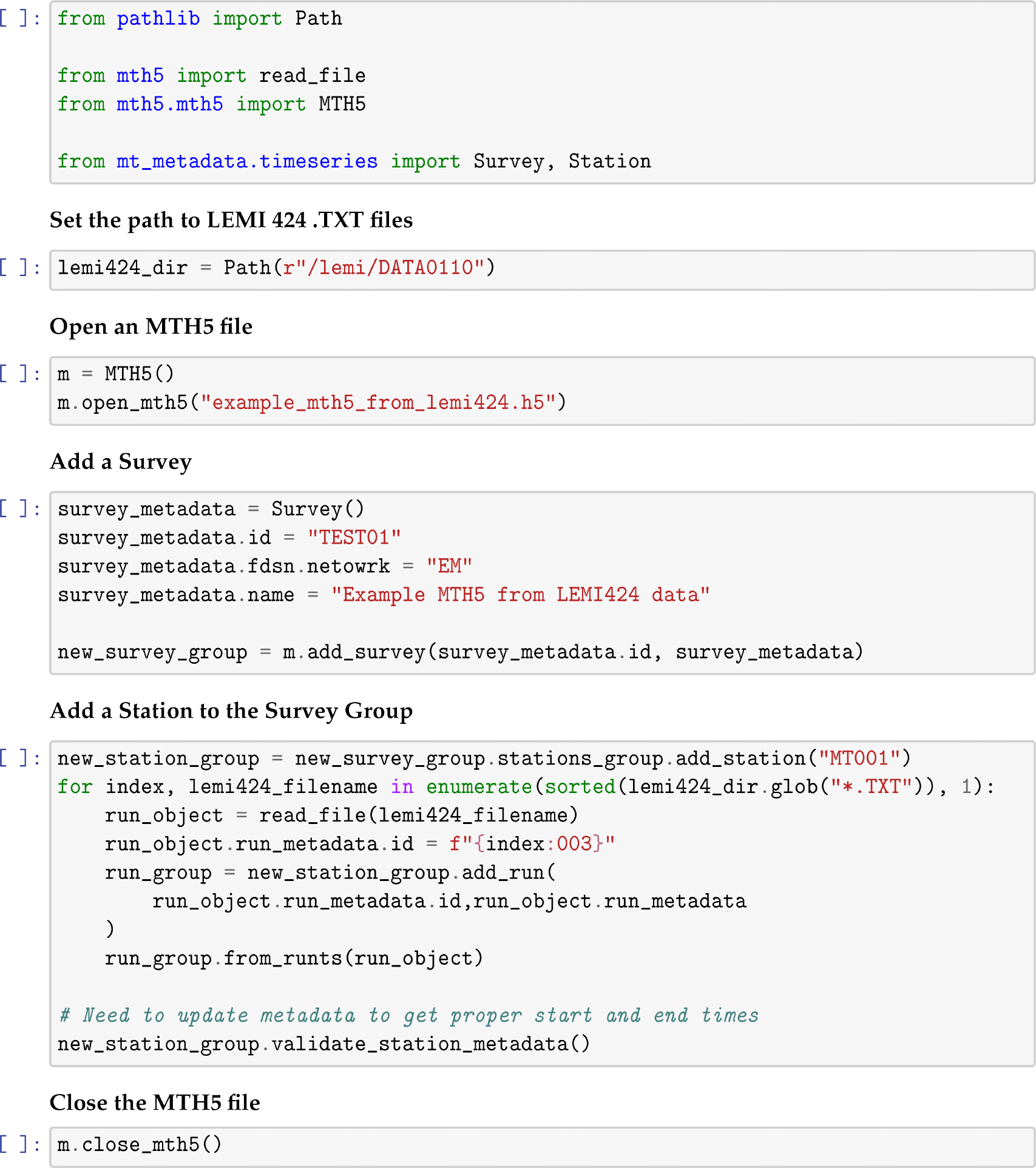}
	\caption{Example code for how to build a MTH5 file from LEMI424 ASCII files.  Note this is a simple example, in a real case much more metadata would need to be input and a new run should only be made if there is a time gap between files. Source code and example LEMI424 files can be found at \url{https://github.com/kujaku11/mth5/tree/master/docs/examples/notebooks/example_01.ipynb}. }
	\label{fig:code02}
\end{figure} 

\section{Example Workflow}

\begin{figure}
	\includegraphics[width=.75\textwidth]{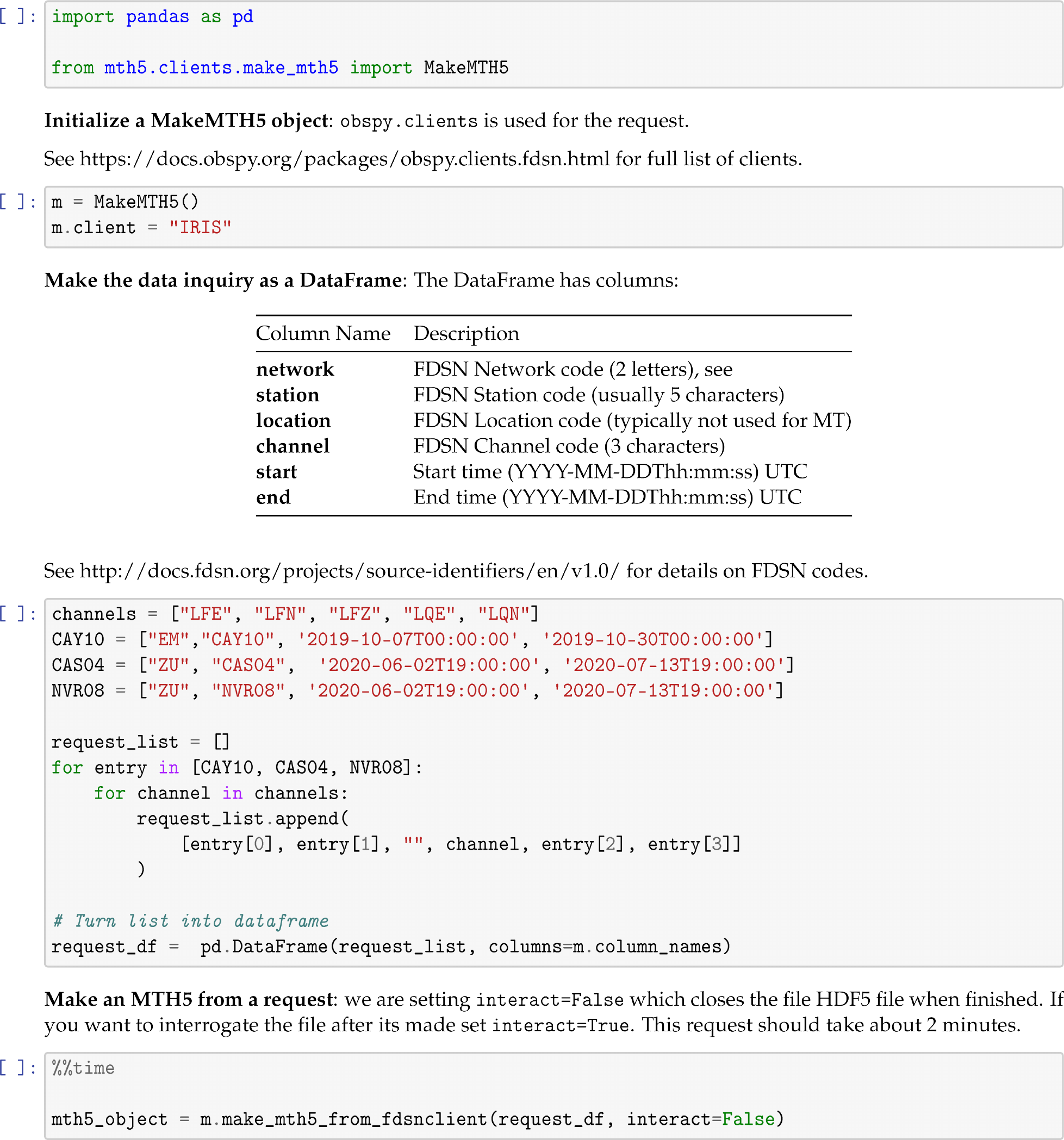}
	\caption{Example code for how to build a MTH5 file from the IRIS DMC.  First, a data/metadata request is made, this can either be a CSV file or a Pandas DataFrame demonstrated here.  If you do not know which data you would like you can use one of many tools provided by IRIS (\url{https://seiscode.iris.washington.edu/}).  The request is done through ObsPy tools where a StationXML (metadata) and miniSEED (data) files are returned.  These are converted into an \texttt{Experiment} (metadata) and \texttt{RunTS} objects (data) to be input into an MTH5 file through the function \texttt{MakeMTH5.make\_mth5\_from\_fdsnclient}. Source code can be found at \url{https://github.com/kujaku11/mth5/blob/master/docs/examples/notebooks/make_mth5_example_iris.ipynb}.}
	\label{fig:code03}
\end{figure}

A MTH5 file is intended to be used both as a format for archiving and as a working format.  Figure \ref{fig:workflow} schematically demonstrates an example workflow intended for data collected with IRIS PASSCAL MT instruments. Two ways to build a MTH5 file exist. First, data from a data logger are directly read into a MTH5 file with accompanying metadata not in the raw file, as described in Section \ref{sec:build} and Figure \ref{fig:code02}.  Second, data are downloaded from the data logger, archived at the IRIS data managment center (DMC), and then retrieved to build a MTH5 file.  This section is focused on the second method (Figure \ref{fig:code03}).

The IRIS DMC provides data in miniSEED (\url{http://www.fdsn.org/pdf/SEEDManual_V2.4.pdf}) and the metadata as a StationXML file (\url{http://www.fdsn.org/xml/station/}). This example uses tools developed in ObsPy for retrieving data and metadata from IRIS DMC. The user should know in advance what data they would like to request. IRIS provides many tools to query the DMC for available data (\url{https://seiscode.iris.washington.edu/}). The data will be represented as an ObsPy \verb|Stream| object, which is a collection of synchronous channels. Tools are built in \verb|mth5.timeseries.RunTS| to read in a \verb|Stream| object. Then the process is similar to Section \ref{sec:build}.       

\section{Suggested Advances to support standardization of MT data and metadata}

The current input/output plug-in architecture in \verb|mth5| only supports a couple data formats.  As the user base grows, more data formats could be added by the community to allow for broader use.  \verb|mt_metadata| could be extended to include transfer functions to standardize reading and writing of various transfer function formats.  Compliance checkers for metadata need to be developed, specifically XSD and JSON schema documents to allow for validation using XML and JSON tools.  An open-source time series processing code is in development, funded through IRIS PASSCAL that uses MTH5 as the working format and \verb|mt_metadata| to output transfer functions.  Existing time series processing codes could be adjusted to work with MTH5 for better cohesion between various processing methods.  A working group managed by the International Association of Geomagnetism and Aeronomy Division VI could be developed to support community driven changes to MT metadata and data standards, which could further support the adherence of MT data to FAIR principles.        

\section{Conclusions}

The MT community has not had a standard format for time series data. Through community development a metadata standard and data format have been developed.  The open-source Python packages \verb|mt_metadata| and \verb|mth5| described in this study provide tools to help standardize MT time series data and make MT time series data more FAIR. Moving forward, MTH5 files will be the working format for data collected with IRIS PASSCAL MT instruments, the standard archive format for data collected by the U.S. Geological Survey, and the standard working format for a time series processing code currently in development by IRIS PASSCAL.  Our release of the MT metadata standard, MTH5 time series data format and the associated open-source packages \verb|mt_metadata| and \verb|mth5| can promote community development of these open-source tools, help standardize MT data in a FAIR way, and provide researchers less familiar with MT time series data with a good entry point to first-hand MT data analysis.

\section{Acknowledgments}

The authors would like to acknowledge the Working Group for Magnetotelluric Data Handling and Software members (Bruce Beaudoin, Lloyd Carothers, Jerry Carter, Gary Egbert, David Goldak, Maeva Pourpoint, Adam Schultz, Maxim Smirnov, Chad Trabant) for defining time series metadata.  The authors would like to acknowledge Ben Murphy for help with fitting pole-zero filters to look-up tables.  This project was partially funded by IRIS and the U.S. Geological Survey Community for Data Integration.  The facilities of the IRIS Consortium are supported by the National Science Foundation’s Seismological Facilities for the Advancement of Geoscience Award under Cooperative Support Agreement EAR-1851048.  Any use of trade, firm, or product names is for descriptive purposes only and does not imply endorsement by the U.S. Government.   

\newpage

\textbf{Code availability section}

\verb|mt_metadata|

Contact: \url{jpeacock@usgs.gov}, 650-439-2833

Hardware requirements: ...

Program language: Python 3.6+
 
Software required: Python 

Program size: 2 Mb

The source codes are available for downloading at the link:
\url{https://doi.org/10.5281/zenodo.5605026}

\verb|mth5|

Contact: \url{jpeacock@usgs.gov}, 650-439-2833

Hardware requirements: ...

Program language: Python 3.6+

Software required: Python 

Program size: 2 Mb

The source codes are available for downloading at the link:
\url{https://doi.org/10.5281/zenodo.5637466}

\bibliographystyle{cas-model2-names}
\bibliography{mth5}

\begin{thebibliography}{28}
\expandafter\ifx\csname natexlab\endcsname\relax\def\natexlab#1{#1}\fi
\providecommand{\url}[1]{\texttt{#1}}
\providecommand{\href}[2]{#2}
\providecommand{\path}[1]{#1}
\providecommand{\DOIprefix}{doi:}
\providecommand{\ArXivprefix}{arXiv:}
\providecommand{\URLprefix}{URL: }
\providecommand{\Pubmedprefix}{pmid:}
\providecommand{\doi}[1]{\href{http://dx.doi.org/#1}{\path{#1}}}
\providecommand{\Pubmed}[1]{\href{pmid:#1}{\path{#1}}}
\providecommand{\bibinfo}[2]{#2}
\ifx\xfnm\relax \def\xfnm[#1]{\unskip,\space#1}\fi
\bibitem[{Bedrosian et~al.(2018)Bedrosian, Peacock, Bowles-Martinez, Schultz
  and Hill}]{Bedrosian2018}
\bibinfo{author}{Bedrosian, P.A.}, \bibinfo{author}{Peacock, J.R.},
  \bibinfo{author}{Bowles-Martinez, E.}, \bibinfo{author}{Schultz, A.},
  \bibinfo{author}{Hill, G.J.}, \bibinfo{year}{2018}.
\newblock \bibinfo{title}{{Crustal inheritance and a top-down control on arc
  magmatism at Mount St Helens}}.
\newblock \bibinfo{journal}{Nature Geoscience} \bibinfo{volume}{11},
  \bibinfo{pages}{865--870}.
\newblock \DOIprefix\doi{10.1038/s41561-018-0217-2}.
\bibitem[{Chave and Jones(2012)}]{Chave2012}
\bibinfo{author}{Chave, A.D.}, \bibinfo{author}{Jones, A.G.},
  \bibinfo{year}{2012}.
\newblock \bibinfo{title}{{The Magnetotelluric Method}}.
\newblock \bibinfo{publisher}{Cambridge University Press}.
\newblock \DOIprefix\doi{10.1017/cbo9781139020138}.
\bibitem[{Chave and Thomson(2004)}]{Chave2004}
\bibinfo{author}{Chave, A.D.}, \bibinfo{author}{Thomson, D.J.},
  \bibinfo{year}{2004}.
\newblock \bibinfo{title}{Bounded influence magnetotelluric response function
  estimation}.
\newblock \bibinfo{journal}{Geophysical Journal International}
  \bibinfo{volume}{157}, \bibinfo{pages}{988--1006}.
\newblock \DOIprefix\doi{10.1111/j.1365-246X.2004.02203.x}.
\bibitem[{Cordell et~al.(2019)Cordell, Unsworth, Diaz, Reyes-Wagner, Currie and
  Hicks}]{Cordell2019}
\bibinfo{author}{Cordell, D.}, \bibinfo{author}{Unsworth, M.J.},
  \bibinfo{author}{Diaz, D.}, \bibinfo{author}{Reyes-Wagner, V.},
  \bibinfo{author}{Currie, C.A.}, \bibinfo{author}{Hicks, S.P.},
  \bibinfo{year}{2019}.
\newblock \bibinfo{title}{{Fluid and melt pathways in the Central Chilean
  subduction zone near the 2010 Maule Earthquake (35--36 S) as inferred from
  magnetotelluric data}}.
\newblock \bibinfo{journal}{Geochemistry, Geophysics, Geosystems}
  \bibinfo{volume}{20}, \bibinfo{pages}{1818--1835}.
\newblock \DOIprefix\doi{10.1029/2018gc008167}.
\bibitem[{Egbert(2002)}]{Egbert2002}
\bibinfo{author}{Egbert, G.D.}, \bibinfo{year}{2002}.
\newblock \bibinfo{title}{Processing and interpretation of electromagnetic
  induction array data}.
\newblock \bibinfo{journal}{Surveys in Geophysics} \bibinfo{volume}{23},
  \bibinfo{pages}{207--249}.
\newblock \DOIprefix\doi{10.1023/A:1015012821040}.
\bibitem[{Habermann et~al.(2021)Habermann, Trabant, Ronan, Bahavar, Crosby,
  Dittman, Carter, Mencin, Suleiman, Carothers and et~al.}]{Habermann2021}
\bibinfo{author}{Habermann, T.}, \bibinfo{author}{Trabant, C.},
  \bibinfo{author}{Ronan, T.}, \bibinfo{author}{Bahavar, M.},
  \bibinfo{author}{Crosby, C.}, \bibinfo{author}{Dittman, T.},
  \bibinfo{author}{Carter, J.}, \bibinfo{author}{Mencin, D.},
  \bibinfo{author}{Suleiman, Y.}, \bibinfo{author}{Carothers, L.},
  \bibinfo{author}{et~al.}, \bibinfo{year}{2021}.
\newblock \bibinfo{title}{{Common data and metadata models for geophysical data
  in the cloud}}.
\newblock \bibinfo{journal}{Earth and Space Science Open Archive} ,
  \bibinfo{pages}{11}\DOIprefix\doi{10.1002/essoar.10509909.1}.
\bibitem[{Harris et~al.(2020)Harris, Millman, van~der Walt, Gommers, Virtanen,
  Cournapeau, Wieser, Taylor, Berg, Smith, Kern, Picus, Hoyer, van Kerkwijk,
  Brett, Haldane, del R{\'{\i}}o, Wiebe, Peterson, G{\'{e}}rard-Marchant,
  Sheppard, Reddy, Weckesser, Abbasi, Gohlke and Oliphant}]{Harris2020}
\bibinfo{author}{Harris, C.R.}, \bibinfo{author}{Millman, K.J.},
  \bibinfo{author}{van~der Walt, S.J.}, \bibinfo{author}{Gommers, R.},
  \bibinfo{author}{Virtanen, P.}, \bibinfo{author}{Cournapeau, D.},
  \bibinfo{author}{Wieser, E.}, \bibinfo{author}{Taylor, J.},
  \bibinfo{author}{Berg, S.}, \bibinfo{author}{Smith, N.J.},
  \bibinfo{author}{Kern, R.}, \bibinfo{author}{Picus, M.},
  \bibinfo{author}{Hoyer, S.}, \bibinfo{author}{van Kerkwijk, M.H.},
  \bibinfo{author}{Brett, M.}, \bibinfo{author}{Haldane, A.},
  \bibinfo{author}{del R{\'{\i}}o, J.F.}, \bibinfo{author}{Wiebe, M.},
  \bibinfo{author}{Peterson, P.}, \bibinfo{author}{G{\'{e}}rard-Marchant, P.},
  \bibinfo{author}{Sheppard, K.}, \bibinfo{author}{Reddy, T.},
  \bibinfo{author}{Weckesser, W.}, \bibinfo{author}{Abbasi, H.},
  \bibinfo{author}{Gohlke, C.}, \bibinfo{author}{Oliphant, T.E.},
  \bibinfo{year}{2020}.
\newblock \bibinfo{title}{{Array programming with {NumPy}}}.
\newblock \bibinfo{journal}{Nature} \bibinfo{volume}{585},
  \bibinfo{pages}{357--362}.
\newblock \DOIprefix\doi{10.1038/s41586-020-2649-2}.
\bibitem[{Heinson et~al.(2018)Heinson, Didana, Soeffky, Thiel and
  Wise}]{Heinson2018}
\bibinfo{author}{Heinson, G.}, \bibinfo{author}{Didana, Y.},
  \bibinfo{author}{Soeffky, P.}, \bibinfo{author}{Thiel, S.},
  \bibinfo{author}{Wise, T.}, \bibinfo{year}{2018}.
\newblock \bibinfo{title}{{The crustal geophysical signature of a world-class
  magmatic mineral system}}.
\newblock \bibinfo{journal}{Scientific Reports} \bibinfo{volume}{8}.
\newblock \DOIprefix\doi{10.1038/s41598-018-29016-2}.
\bibitem[{Hoyer et~al.(2021)Hoyer, Hamman, Roos, {Keewis}, Cherian, Fitzgerald,
  Fujii, Maussion, Hauser, {Crusaderky}, Clark, Kleeman, Kluyver, Munroe,
  Amici, Nicholas, Barghini, Banihirwe, {Gimperiale}, {Zac Hatfield-Dodds},
  Abernathey, {Illviljan}, Bell, {Johnomotani}, Roszko, Wolfram, Signell,
  Mühlbauer, Sinai and {Benoit Bovy}}]{Hoyer2021}
\bibinfo{author}{Hoyer, S.}, \bibinfo{author}{Hamman, J.},
  \bibinfo{author}{Roos, M.}, \bibinfo{author}{{Keewis}},
  \bibinfo{author}{Cherian, D.}, \bibinfo{author}{Fitzgerald, C.},
  \bibinfo{author}{Fujii, K.}, \bibinfo{author}{Maussion, F.},
  \bibinfo{author}{Hauser, M.}, \bibinfo{author}{{Crusaderky}},
  \bibinfo{author}{Clark, S.}, \bibinfo{author}{Kleeman, A.},
  \bibinfo{author}{Kluyver, T.}, \bibinfo{author}{Munroe, J.},
  \bibinfo{author}{Amici, A.}, \bibinfo{author}{Nicholas, T.},
  \bibinfo{author}{Barghini, A.}, \bibinfo{author}{Banihirwe, A.},
  \bibinfo{author}{{Gimperiale}}, \bibinfo{author}{{Zac Hatfield-Dodds}},
  \bibinfo{author}{Abernathey, R.}, \bibinfo{author}{{Illviljan}},
  \bibinfo{author}{Bell, R.}, \bibinfo{author}{{Johnomotani}},
  \bibinfo{author}{Roszko, M.}, \bibinfo{author}{Wolfram, P.J.},
  \bibinfo{author}{Signell, J.}, \bibinfo{author}{Mühlbauer, K.},
  \bibinfo{author}{Sinai, Y.B.}, \bibinfo{author}{{Benoit Bovy}},
  \bibinfo{year}{2021}.
\newblock \bibinfo{title}{{pydata/xarray: v0.18.2}}.
\newblock \DOIprefix\doi{10.5281/ZENODO.4774304}.
\bibitem[{Hoyer and Hamman(2017)}]{Hoyer2017}
\bibinfo{author}{Hoyer, S.}, \bibinfo{author}{Hamman, J.J.},
  \bibinfo{year}{2017}.
\newblock \bibinfo{title}{{xarray: N-D labeled Arrays and Datasets in Python}}.
\newblock \bibinfo{journal}{Journal of Open Research Software}
  \bibinfo{volume}{5}.
\newblock \DOIprefix\doi{10.5334/jors.148}.
\bibitem[{Kara{\c{s}} et~al.(2020)Kara{\c{s}}, Tank, Ogawa, Oshiman, Matsushima
  and Honkura}]{Karas2020}
\bibinfo{author}{Kara{\c{s}}, M.}, \bibinfo{author}{Tank, S.B.},
  \bibinfo{author}{Ogawa, Y.}, \bibinfo{author}{Oshiman, N.},
  \bibinfo{author}{Matsushima, M.}, \bibinfo{author}{Honkura, Y.},
  \bibinfo{year}{2020}.
\newblock \bibinfo{title}{{Probing the relationship between electrical
  conductivity and creep through upper crustal fluids along the western part of
  the North Anatolian Fault with three-dimensional magnetotellurics}}.
\newblock \bibinfo{journal}{Tectonophysics} \bibinfo{volume}{791},
  \bibinfo{pages}{228561}.
\newblock \DOIprefix\doi{10.1016/j.tecto.2020.228561}.
\bibitem[{Kelbert(2019)}]{Kelbert2019}
\bibinfo{author}{Kelbert, A.}, \bibinfo{year}{2019}.
\newblock \bibinfo{title}{{The Role of Global/Regional Earth Conductivity
  Models in Natural Geomagnetic Hazard Mitigation}}.
\newblock \bibinfo{journal}{Surveys in Geophysics} \bibinfo{volume}{41},
  \bibinfo{pages}{115--166}.
\newblock \DOIprefix\doi{10.1007/s10712-019-09579-z}.
\bibitem[{Kelbert(2020)}]{Kelbert2020}
\bibinfo{author}{Kelbert, A.}, \bibinfo{year}{2020}.
\newblock \bibinfo{title}{{EMTF XML: New data interchange format and conversion
  tools for electromagnetic transfer functions}}.
\newblock \bibinfo{journal}{{Geophysics}} \bibinfo{volume}{85},
  \bibinfo{pages}{F1--F17}.
\newblock \DOIprefix\doi{10.1190/geo2018-0679.1}.
\bibitem[{Kelbert et~al.(2011)Kelbert, Egbert and Schultz}]{Kelbert2011}
\bibinfo{author}{Kelbert, A.}, \bibinfo{author}{Egbert, G.D.},
  \bibinfo{author}{Schultz, A.}, \bibinfo{year}{2011}.
\newblock \bibinfo{title}{{Data Services Products: EMTF, The Magnetotelluric
  Transfer Functions}}.
\newblock \DOIprefix\doi{10.17611/DP/EMTF.1}.
\bibitem[{Kelbert et~al.(2018)Kelbert, Erofeeva, Trabant, Karstens and
  Fossen}]{Kelbert2018}
\bibinfo{author}{Kelbert, A.}, \bibinfo{author}{Erofeeva, S.},
  \bibinfo{author}{Trabant, C.}, \bibinfo{author}{Karstens, R.},
  \bibinfo{author}{Fossen, M.V.}, \bibinfo{year}{2018}.
\newblock \bibinfo{title}{{Taking Magnetotelluric Data out of the Drawer}}.
\newblock \bibinfo{journal}{EOS} \bibinfo{volume}{99}.
\newblock \DOIprefix\doi{10.1029/2018eo112859}.
\bibitem[{Klein and Taaheri(2016)}]{Klein2016}
\bibinfo{author}{Klein, L.}, \bibinfo{author}{Taaheri, A.},
  \bibinfo{year}{2016}.
\newblock \bibinfo{title}{{HDF-EOS5 Data Model, File Format and Library}}.
\newblock \bibinfo{type}{Tech Report} \bibinfo{number}{ESDS-RFC-008v1.0}.
  National Aeronautics and Space Administration.
\bibitem[{Krischer et~al.(2015)Krischer, Megies, Barsch, Beyreuther, Lecocq,
  Caudron and Wassermann}]{Krischer2015}
\bibinfo{author}{Krischer, L.}, \bibinfo{author}{Megies, T.},
  \bibinfo{author}{Barsch, R.}, \bibinfo{author}{Beyreuther, M.},
  \bibinfo{author}{Lecocq, T.}, \bibinfo{author}{Caudron, C.},
  \bibinfo{author}{Wassermann, J.}, \bibinfo{year}{2015}.
\newblock \bibinfo{title}{{ObsPy: a bridge for seismology into the scientific
  Python ecosystem}}.
\newblock \bibinfo{journal}{Computational Science \& Discovery}
  \bibinfo{volume}{8}, \bibinfo{pages}{014003}.
\newblock \DOIprefix\doi{10.1088/1749-4699/8/1/014003}.
\bibitem[{Krischer et~al.(2016)Krischer, Smith, Lei, Lefebvre, Ruan,
  de~Andrade, Podhorszki, Bozda{\u{g}} and Tromp}]{Krischer2016}
\bibinfo{author}{Krischer, L.}, \bibinfo{author}{Smith, J.},
  \bibinfo{author}{Lei, W.}, \bibinfo{author}{Lefebvre, M.},
  \bibinfo{author}{Ruan, Y.}, \bibinfo{author}{de~Andrade, E.S.},
  \bibinfo{author}{Podhorszki, N.}, \bibinfo{author}{Bozda{\u{g}}, E.},
  \bibinfo{author}{Tromp, J.}, \bibinfo{year}{2016}.
\newblock \bibinfo{title}{{An Adaptable Seismic Data Format}}.
\newblock \bibinfo{journal}{Geophysical Journal International}
  \bibinfo{volume}{207}, \bibinfo{pages}{1003--1011}.
\newblock \DOIprefix\doi{10.1093/gji/ggw319}.
\bibitem[{Lee(2012)}]{Lee2012}
\bibinfo{author}{Lee, J.}, \bibinfo{year}{2012}.
\newblock \bibinfo{title}{{GLAS\_HDF detailed design}}.
\newblock \bibinfo{type}{Tech Report}. National Aeronautics and Space
  Administration.
\bibitem[{Loomis et~al.(2019)Loomis, Luthcke and Sabaka}]{Loomis2019}
\bibinfo{author}{Loomis, B.D.}, \bibinfo{author}{Luthcke, S.B.},
  \bibinfo{author}{Sabaka, T.J.}, \bibinfo{year}{2019}.
\newblock \bibinfo{title}{{Regularization and error characterization of GRACE
  mascons}}.
\newblock \bibinfo{journal}{Journal of Geodesy} \bibinfo{volume}{93},
  \bibinfo{pages}{1381--1398}.
\newblock \DOIprefix\doi{10.1007/s00190-019-01252-y}.
\bibitem[{McKinney(2010)}]{McKinney2010}
\bibinfo{author}{McKinney, W.}, \bibinfo{year}{2010}.
\newblock \bibinfo{title}{{Data structures for statistical computing in
  Python}}, in: \bibinfo{booktitle}{Proceedings of the 9th Python in Science
  Conference}, \bibinfo{publisher}{{SciPy}}.
\newblock \DOIprefix\doi{10.25080/majora-92bf1922-00a}.
\bibitem[{Murphy et~al.(2021)Murphy, Lucas, Love, Kelbert, Bedrosian and
  Rigler}]{Murphy2021}
\bibinfo{author}{Murphy, B.S.}, \bibinfo{author}{Lucas, G.M.},
  \bibinfo{author}{Love, J.J.}, \bibinfo{author}{Kelbert, A.},
  \bibinfo{author}{Bedrosian, P.A.}, \bibinfo{author}{Rigler, E.J.},
  \bibinfo{year}{2021}.
\newblock \bibinfo{title}{{Magnetotelluric sampling and geoelectric hazard
  estimation: are national-scale surveys sufficient?}}
\newblock \bibinfo{journal}{Space Weather} \bibinfo{volume}{19}.
\newblock \DOIprefix\doi{10.1029/2020sw002693}.
\bibitem[{{Open Geospatial Consortium}(2019)}]{OGC2019}
\bibinfo{author}{{Open Geospatial Consortium}}, \bibinfo{year}{2019}.
\newblock \bibinfo{title}{{Geographic information — well-known text
  representation of coordinate reference systems}}.
\newblock \bibinfo{type}{Technical Report}. {Open Geospatial Consortium, eds R.
  Lott}.
\newblock \URLprefix
  \url{http://docs.opengeospatial.org/is/18-010r7/18-010r7.html#8}.
\bibitem[{Peacock et~al.(2020)Peacock, Earney, Mangan, Schermerhorn, Glen,
  Walters and Hartline}]{Peacock2020}
\bibinfo{author}{Peacock, J.R.}, \bibinfo{author}{Earney, T.E.},
  \bibinfo{author}{Mangan, M.T.}, \bibinfo{author}{Schermerhorn, W.D.},
  \bibinfo{author}{Glen, J.M.}, \bibinfo{author}{Walters, M.},
  \bibinfo{author}{Hartline, C.}, \bibinfo{year}{2020}.
\newblock \bibinfo{title}{{Geophysical characterization of the Northwest
  Geysers geothermal field, California}}.
\newblock \bibinfo{journal}{Journal of Volcanology and Geothermal Research}
  \bibinfo{volume}{399}, \bibinfo{pages}{106882}.
\newblock \DOIprefix\doi{10.1016/j.jvolgeores.2020.106882}.
\bibitem[{Peacock et~al.(2021)Peacock, Frassetto, Kelbert, Egbert, Smirnov,
  Schultz, Kappler, Ronan and Trabant}]{Peacock2021}
\bibinfo{author}{Peacock, J.R.}, \bibinfo{author}{Frassetto, A.},
  \bibinfo{author}{Kelbert, A.}, \bibinfo{author}{Egbert, G.D.},
  \bibinfo{author}{Smirnov, M.}, \bibinfo{author}{Schultz, A.},
  \bibinfo{author}{Kappler, K.}, \bibinfo{author}{Ronan, T.},
  \bibinfo{author}{Trabant, C.}, \bibinfo{year}{2021}.
\newblock \bibinfo{title}{{Metadata standards for magnetotelluric time series
  data}}.
\newblock \bibinfo{howpublished}{U.S. Geological Survey data release}.
\newblock \DOIprefix\doi{https://doi.org/10.5066/P9AXGKEV}.
\bibitem[{Reback et~al.(2021)Reback, {Jbrockmendel}, McKinney, Van Den~Bossche,
  Augspurger, Cloud, Hawkins, Roeschke, {Gfyoung}, {Sinhrks}, Klein, Hoefler,
  {Terji Petersen}, Tratner, She, Ayd, Naveh, {JHM Darbyshire}, Garcia,
  Shadrach, Schendel, Hayden, Saxton, Gorelli, {Fangchen Li}, Zeitlin,
  Jancauskas, McMaster, Battiston and {Skipper Seabold}}]{Reback2021}
\bibinfo{author}{Reback, J.}, \bibinfo{author}{{Jbrockmendel}},
  \bibinfo{author}{McKinney, W.}, \bibinfo{author}{Van Den~Bossche, J.},
  \bibinfo{author}{Augspurger, T.}, \bibinfo{author}{Cloud, P.},
  \bibinfo{author}{Hawkins, S.}, \bibinfo{author}{Roeschke, M.},
  \bibinfo{author}{{Gfyoung}}, \bibinfo{author}{{Sinhrks}},
  \bibinfo{author}{Klein, A.}, \bibinfo{author}{Hoefler, P.},
  \bibinfo{author}{{Terji Petersen}}, \bibinfo{author}{Tratner, J.},
  \bibinfo{author}{She, C.}, \bibinfo{author}{Ayd, W.}, \bibinfo{author}{Naveh,
  S.}, \bibinfo{author}{{JHM Darbyshire}}, \bibinfo{author}{Garcia, M.},
  \bibinfo{author}{Shadrach, R.}, \bibinfo{author}{Schendel, J.},
  \bibinfo{author}{Hayden, A.}, \bibinfo{author}{Saxton, D.},
  \bibinfo{author}{Gorelli, M.E.}, \bibinfo{author}{{Fangchen Li}},
  \bibinfo{author}{Zeitlin, M.}, \bibinfo{author}{Jancauskas, V.},
  \bibinfo{author}{McMaster, A.}, \bibinfo{author}{Battiston, P.},
  \bibinfo{author}{{Skipper Seabold}}, \bibinfo{year}{2021}.
\newblock \bibinfo{title}{{pandas-dev/pandas: Pandas 1.3.0}}.
\newblock \DOIprefix\doi{10.5281/ZENODO.3509134}.
\bibitem[{{The HDF Group}(2019)}]{HDFGroup2019}
\bibinfo{author}{{The HDF Group}}, \bibinfo{year}{2019}.
\newblock \bibinfo{title}{{HDF5 Users's Guide: HDF5 Release 1.10}}.
\newblock \bibinfo{type}{Technical Report}. {The HDF Group}.
\newblock \URLprefix \url{https://portal.hdfgroup.org/display/HDF5/HDF5 User
  Guides?preview=/53610087/53610088/Users_Guide.pdf}.
\bibitem[{Wilkinson et~al.(2016)Wilkinson, Dumontier and I.~J.
  J.~Aalbersberg}]{Wilkinson2016}
\bibinfo{author}{Wilkinson, M.D.}, \bibinfo{author}{Dumontier, M.},
  \bibinfo{author}{I.~J. J.~Aalbersberg, e.}, \bibinfo{year}{2016}.
\newblock \bibinfo{title}{{The {FAIR} Guiding Principles for scientific data
  management and stewardship}}.
\newblock \bibinfo{journal}{Scientific Data} \bibinfo{volume}{3}.
\newblock \DOIprefix\doi{10.1038/sdata.2016.18}.

\end{thebibliography}

\end{document}